\documentclass{article}    
\pdfoutput=1

\usepackage{longtable}
\usepackage{graphicx}

\begin{document}                                                                                   
\centerline{\bf{Morphologies of AGN host galaxies using HST/ACS }}
\centerline{\bf{in the CDFS-GOODS field}} 
\centerline{\bf{Priya Hasan}} 
\noindent$^1$Inter-University Centre for Astronomy and Astrophysics, Ganeshkhind,\\ Pune~-~411007,~India. 
\newline 
\centerline{and}
\newline $^2$Department of Astronomy, Osmania University, \\ Hyderabad~-~500007,~India.
\date{\today}

\begin{abstract}
Using HST/ACS images in four bands F435W, F606W, F775W and F850LP, we identify optical counterparts to the X-ray sources in the Chandra Deep Field South in the GOODS South field. A detailed study has been made of these sources to study their morphological types. We use methods like decomposition of galaxy luminosity profiles, color maps and visual inspection of 192 galaxies which are identified as possible optical counterparts of Chandra X-ray sources in the CDFS-GOODS field. We find that most moderate luminosity AGN hosts are bulge dominated in the redshift range ($z \approx 0.4-1.3$), but not merging/interacting galaxies. This implies probable fueling of the moderate luminosity AGN by mechanisms other than those merger driven.
\end{abstract}
{Galaxies, AGN, X-ray emission, Morphology}

\section{Introduction}
Many clues to galaxy formation lay hidden in the fine details of galaxy structure. Active galactic nuclei (AGN) are intimately related to formation of bulges and black holes. A key question in astrophysics is the nature of the AGN host galaxies. Some of the recent developments in this area include the discovery that most nearby massive galaxies harbour central supermassive black holes (SMBHs; Magorrian et al. 1998), that AGNs  the local universe reside predominantly in bulge-dominated host galaxies (Kauffman et al. 2003), and that a tight correlation exists locally between SMBH mass and host galaxy properties such as bulge velocity dispersion and light-profile concentration (Ferrarese \& Merritt 2000; Gebhardt et al. 2000; Graham et al. 2001). The Great Observatories Origins Deep Survey (GOODS) provides a homogenous sample of deep, multiwavelength observations of galaxies using data from Hubble, Spitzer Space Telescope, Chandra and XMM-Newton (Giavalisco et al 2004).  This provides a very unique database of AGNs, starburst galaxies and normal galaxies in wavelengths ranging from infrared to X-ray.

                        The GOODS database can be used to
understand the morphology, spectral properties and evolution process of
peculiar objects (with some indication of AGN activity) in this sample
using indicators like their X-ray luminosity, radio flux, etc. Grogin et al (2005) have studied this field using CAS Conselice parameters (Conselice 2003), and concluded to no merger-AGN connection for moderate luminosity AGN for the same field. We would like to study the morphology of these galaxies using a different method to examine the above result in greater detail.

This work includes cross correlation of  X-ray source catalogues (Giacconi et al 2002) and the HST/ACS $z_{850}$ catalogs  of the GOODS field obtained with
Chandra Deep Field South (CDFS) survey with catalogues in the optical. Refined techniques have been used to identify most probable counterparts. 

Morphological classification is essential to reveal the nature of AGN host galaxies. However, at high redshifts, it becomes difficult to classify galaxy morphology securely because the images of high redshift galaxies suffer from reduced resolution, bandshifting and cosmological surface brightness dimming effects, compared with the local objects. With HST ACS, high resolution imaging in two or more bands with spatially resolved color distribution can be used to investigate the distribution of the stellar population, which is complementary to single band images. Star forming regions and dusty regions also stand out very well in color maps. 

We have carried out decomposition of the luminosity profiles of the galaxies using GALFIT (Peng et al 2002), to derive the bulge to disk ratios and possible features in residual maps.

X-ray colors are good indicators of population types in galaxies. This data is available in Alexander et al 2003 and the CDFS 1 M
sec catalogue, we derive the counts in three different energy
bins soft(0.3-1 keV), medium (1-2 keV) and hard (2-8keV).

This paper is organised as follows. Section~1 is the introduction and explains the significance of the present work. Section~2 describes the HST ACS observations and the methods we adopt for identifying counterparts.  Section~3 describes the working of GALFIT which is a two-dimensional program used for the decomposition of the luminosity profiles of galaxies to obtain the structural parameters. Section~4 describes the method to produce color maps and lists them. Section 5 and 6 describe the results and conclusions. Throughout this paper, we adopt $H_{0}=$ 70 km s${-1}$ Mpc${-1}$, $\Omega_{M}= 0.3$ and  $\Omega_{\Lambda}=0.7$.


\section{Observations and Methods for identification of counterparts}
The publicly available version v1.0 of the reduced, calibrated, stacked and mosaiced images of the GOODS-South fields, which forms a subset of the CDFS field in four bands, namely F435W, F606W, F775W and F850LP i.e. $B$, $V$, $i$ and $z$. This was coupled with the catalogues (1) publicly available SExtractor (Bertin and Arnouts 1996) based version 1.1 of the ACS multiband source catalogue released by the GOODS  team. (2) X-ray source catalogues of the CDFS (Giacconi et al 2002).

The observed AGN sky density in the Chandra Deep Fields is $\approx$ 7200 deg$^{-2}$ (Bauer et al 2004). This exceptional effectiveness at finding AGN arises largely because X-ray selection (1) has reduced absorption bias (2) has minimal dilution by host-galaxy starlight and (3) allows concentration of intensive optical spectroscopic follow-up upon high probability AGN with faint optical counterparts. The AGN sky density from the Chandra Deep Fields exceeds that found at any other wavelength and is 10-20 times higher than that found in the deepest optical spectroscopic surveys. (Wolf et al 2003, Hunt et al 2004); only optical variability studies (Sarajdini, Gilliland \& Kasm 2003) may be generating comparable AGN sky densities.
Since we have multiband data, we have used rest-frame B-band magnitudes by using images in the appropriate wavebands. This is done to be on the safer side since kcorrection for each component (bulge, disc, etc) maybe different hence leading to unwanted and unaccessible errors creeping up.

\begin{figure} 
\centerline{\includegraphics[width=20pc,height=20pc]{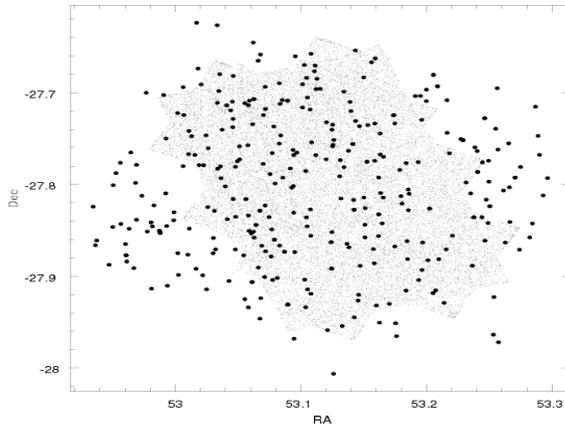}}
\caption[]{Overlap between the GOODS \& CDFS fields}
\label{gc}
\end{figure}

Figure \ref{gc} shows the overlap between the GOODS \& CDFS fields. There are about 230 Chandra sources contained in the GOODS field. In the search for optical counterparts of Chandra X ray sources, care has to be taken because of the difference in resolution of the telescopes. The majority of authors use the closest match as the `most probable' or best match. We present two methods for identification of most probable matches.

Another caveat is that many of the Xray selected Chandra sources are a  mixture of AGN and  comparatively nearby and optically bright starbursts and `quiescent' galaxies. At redshifts larger than 0.4 the contamination is negligible and hence to be sure that our sample contains only AGN, we choose $z=0.4$ as our lower limit for redshift evolution analysis (Grogin et al 2005). 


\begin{figure*}
\centerline{\includegraphics[width=20pc,height=20pc]{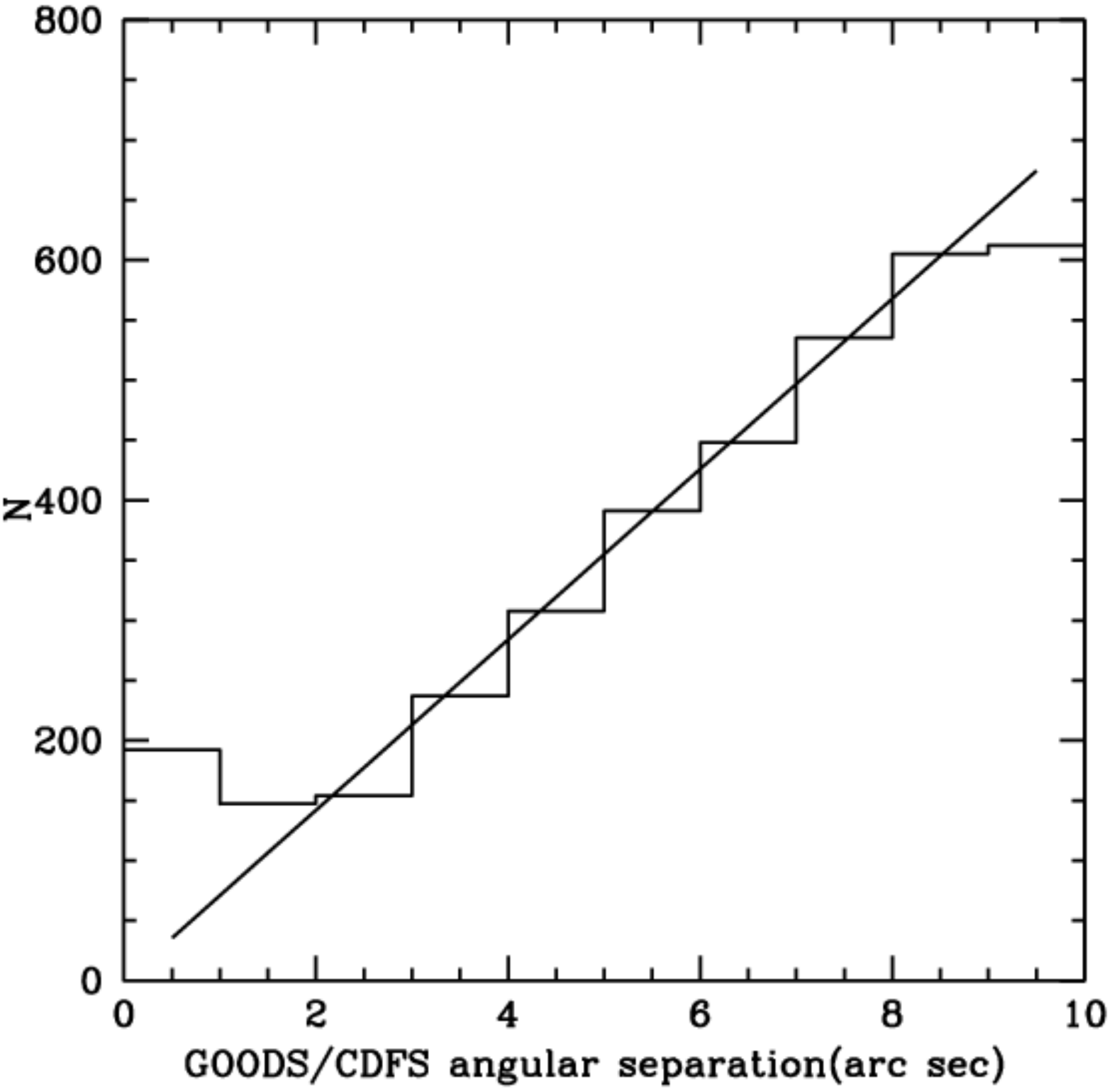}}
\caption{The increase in number of sources within annuli $1^{"}$ wide of increasing radius. Also shown is the corresponding increase for a uniform spatial distribution.}
\label{wad}
\end{figure*}

The first method for identification of counterparts was the one adopted by Wadadekar \& Kembhavi (1999) shown in Fig. \ref{wad}. The total number of GOODS sources is 29,599 for an area of 160 sq arc min. We find the average density of sources that can fall within a search radius drawn around each Chandra source. For the actual data, we increase the search radius in steps of $1^{"}$ around each Chandra source and plot the angular separation versus the increase in number of matches. As seen in the Fig. \ref{wad}, the number of matches increases linearly with $r$ for a uniform distribution and a clear peak is noticed around a radius of $2^{"}$ .

\begin{figure*}
\centerline{\includegraphics[width=20pc,height=20pc]{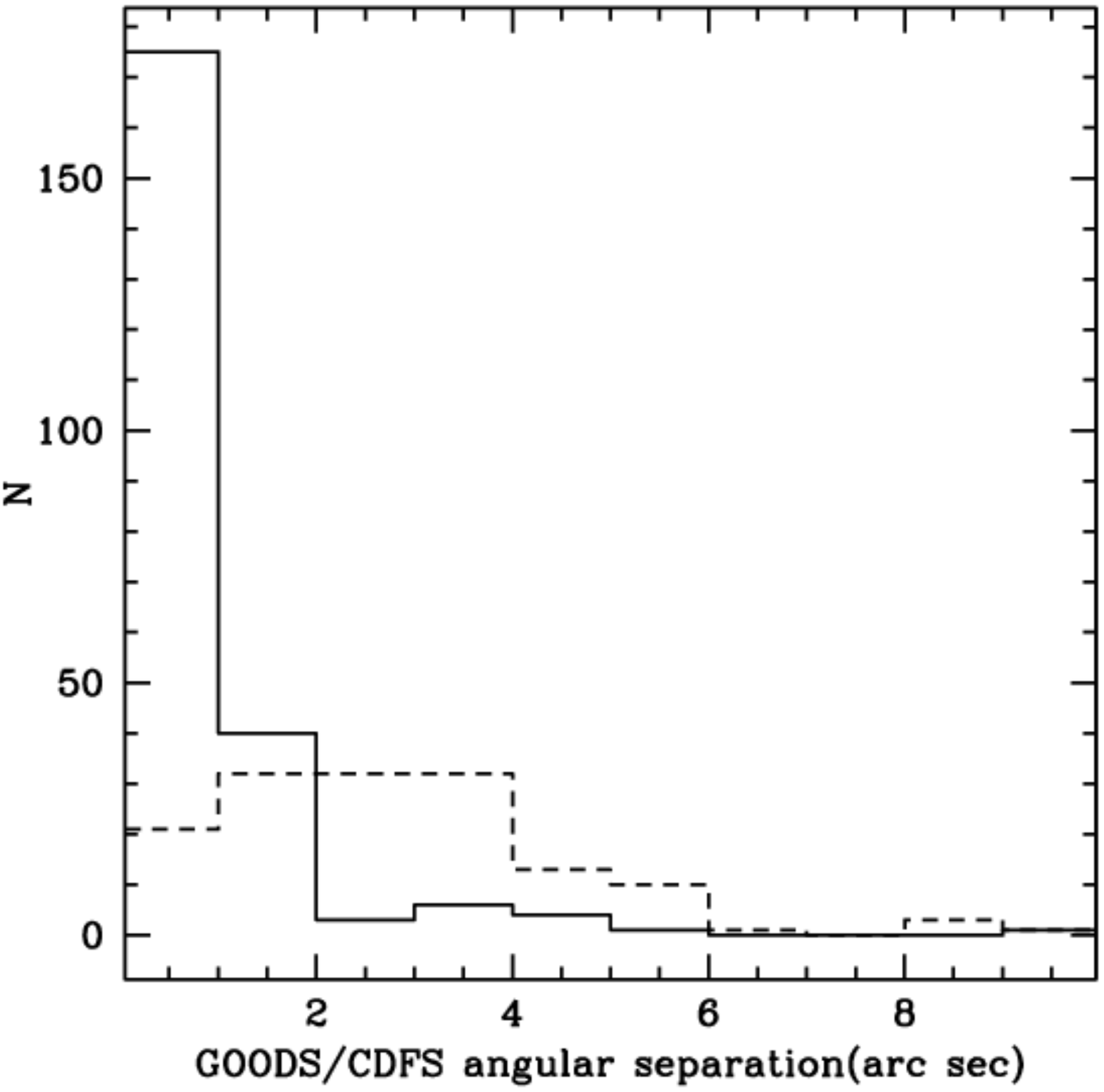}}
\caption{The angular separation between closest counterparts in the GOODS and CDFS sources, for the data (solid line) simulated data obtained by shifting one catalogue by $0.1^{0}$ (dashed line)}
\label{bis}
\end{figure*}

Another method adopted was the one by Bischof \& Becker (1997).
Here we find the nearest optical sources of the GOODS catalogue to the CDFS point source catalogue (Giacconi, 2002). We estimate the chance coincidence rate by shifting the catalogue by a small amount $0.1^{0}$ and then making a similar plot. Figure \ref{bis} shows the results of this, thus showing that with a radius of $2.0^{"}$ the probability of a chance coincidence is very small. Such a method retains the structure of the field as we do not assume any distribution function.

\begin{figure*}
\centerline{\includegraphics[width=20pc,height=20pc]{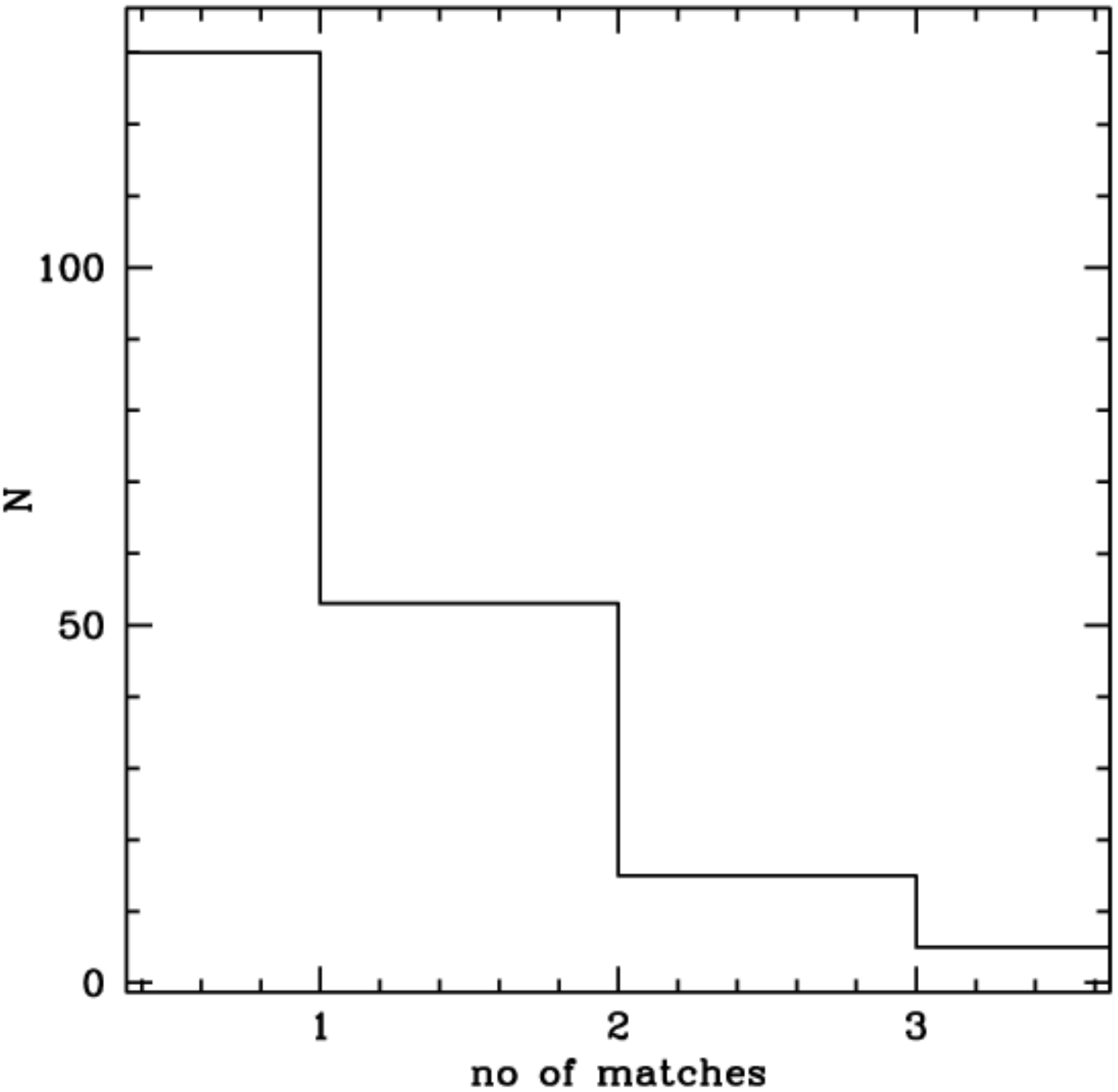}}
\caption{Distribution of the angular separations of the `best' match to CDFS sources}
\label{matchno}
\end{figure*}

There are 203 Chandra sources contained in the GOODS area.
Based on our earlier discussion we had used a $2.0^{"}$ search radius to look for optical counterparts of the Chandra sources.
Within that radius there were multiple matches found. Figure \ref{matchno} shows the number of matches found within a $2^{"}$ radius of the Chandra source.

The majority of authors use the closest match as the `most probable' or best match. If it was found that the brightest match was the closest one, that was adopted as the most likely match (164 matches). Ambiguity was only when the closest match was fainter \& there was a brighter match further away. In the such cases (19), a comparison was made of the probability of the chance occurrence of a source of $that$ magnitude at $that$ particular distance. The source that had the lower probability of a chance occurance was adopted as the likely match. In the case of 4 such pairs, even these comparisons were not sufficient to identify the likely counterpart.

Therefore, we have 164 (one counterpart) + 15 + 4 (ambiguous) counterparts. There were still 20 sources without counterparts in $2^{"}$.

In our method of identification, we have a quantitative measure of the `likely match' hence making it a more reliable method. The number of counterparts in all these methods are similar at the brighter end, but at the fainter end this method is more reliable, as it determines matches as a function of brightness as well as proximity.

\subsection{Redshifts}
Spectroscopy is essential to achieve the scientific goals of GOODS and provides the time coordinate required to understand the evolution of galaxy masses, morphologies, clustering and various aspects. As part of the Great Observatories Origins Deep Survey (GOODS), multi-object, optical to NIR spectroscopy in the Chandra Deep Field South (CDFS) has been carried out, using the FORS2 instrument mounted at the Kueyen Unit Telescope of the VLT at ESO's Cerro Paranal Observatory, Chile.

The spectroscopic data used is from the GOODS FORS2 Spectroscopy Release: Version 2.0. This release contains the results of the ESO/GOODS large program LP170.A-0788 (PI Cesarsky) of spectroscopy of faint galaxies in the Chandra Deep Field South (CDFS). 1204 spectra of 930 unique targets have been obtained in service mode with the FORS2 spectrograph during the period September 2002 - February 2004, providing in total 943 redshift measurements with quality flag A, B or C (A=solid redshift, B=likely redshift, C=potential redshift). 725 of the unique targets have an assigned redshift with quality flag A, B or C. The average of the redshift distribution is z=1.5 (median z=1.1). The typical redshift uncertainty is estimated to be +/-0.005. Spectroscopic redshifts are publicly available from ESO/GOODS project (Vanzella et al., 2005,2006). Other than this data from the CDFS Master Spectroscopic Catalog - v1.2 has also been used.

Based on the photometry of 10 near-UV, optical, and near infrared bands of the CDFS with observations from ESO and the HST , photometric redshifts are available for 342 Xray sources, which constitute ~ 99 \% of all the detected Xray sources in the field (Zheng et al, 2004).

\section{Morphologies of the AGN sample}
Based on the Hubble ACS imaging, two dimensional fitting has been done to derive structural parameters using GALFIT (Peng et al 2002). With the structural parameters and information provided from color maps, the morphological classification has been performed.
\subsection{Structural parameters}

The two dimensional decomposition has been performed using GALFIT. In order to model galaxy profiles with a maximum degree of flexibility, GALFIT uses a number of functions and can combine an arbitrary number of them simultaneously. GALFIT uses convolution, but it can be turned off if not needed. Convolution is done by using the convolution theorem where it multiplies the Fourier transforms of the PSF and the models, and then inverse transforms them.

     The four input images for GALFIT are the CCD image of the galaxy, a noise array (in this case the sigma maps), a PSF, and an optional dust (or bad pixel) maskall in FITS image file format. Pixels in the dust mask are rejected from the fit.

     During the fit (the reduced $\chi^{2}$) is minimized, defined in the standard way as

$$\chi^{2}_{\nu}=\frac{1}{N_{dof}} \sum_{x=1}^{nx} \sum_{y=1}^{nx} \frac{(flux_{x,y}-model_{x,y})^{2}}{\sigma^{2}_{x,y}}$$

where

$$model_{x,y} = \sum_{\nu =1}^{nf} f_{\nu,x,y}(\alpha_{1}...\alpha_{n})$$

$N_{dof}$ is the number of degrees of freedom in the fit; $nx$ and $ny$ are the $x$ and $y$ image dimensions; and $flux_{x,y}$ is the image flux at pixel (x, y). The $model_{x,y}$ is the sum of the $nf$ functions $f_{\nu,x,y}(\alpha_{1},...,\alpha_{n})$ employed, where {1,...,n} are the two-dimensional model parameters. The uncertainty as a function of pixel position $\sigma_{x,y}$ is the Poisson error at each pixel, which can be provided as an input image.

 The Sersic profile has the following form:
$$\Sigma(r)= \Sigma_{e} e^{-k[(r/r_{e})^{1/n} - 1]}$$

where $r_{e}$ is the effective radius of the galaxy, $e$ is the surface brightness at $r_{e}$, $n$ is the power-law index, and $k$ is coupled to $n$ such that half of the total flux is always within $r_{e}$. For $n\geq 2$, $k \approx 2n - 0.331$; at low $n$, $k(n)$ flattens out toward 0 and is obtained by interpolation. The original de Vaucouleurs (1948) profile is a special case with $n$ = 4 and $k= 7.67$. While the de Vaucouleurs profile is well suited for `classical' bulges, some bulges may be better represented by exponential profiles (e.g., Kormendy \& Bruzual 1978; Shaw \& Gilmore 1989; Kent, Dame, \& Fazio 1991; Andrekakis \& Sanders 1994). The elegance of the Sersic profile is that it forms a continuous sequence from a Gaussian ($n$ = 0.5) to an exponential ($n$ = 1) to a de Vaucouleurs ($n$ = 4) profile simply by varying the exponent. It is very useful for modeling bars and flat disks; the smaller the index $n$ is, the faster the core flattens within $r < r_{e}$, and the steeper the intensity drop beyond $r > r_{e}$. The flux, integrated over all radii for an elliptical Sersic profile with an axis ratio $q$, is

$$F_{tot}=2\pi r_{e}^{2}\Sigma_{e} e^{k}nk^{-2n}\Gamma(2n)q/R(c)$$

where $\Gamma(2n)$ is the gamma function. $F_{tot}$ is converted into a magnitude by GALFIT using the standard FITS exposure time parameter (EXPTIME) in the image header. $R(c)$ is a function that accounts for the area ratio between a perfect ellipse and a generalized ellipse of diskiness/boxiness parameter $c$, given by

$$R(c)=\frac{\pi c}{4\beta (1/c,1+1/c)}$$

where $\beta(1/c,1 + 1/c)$ is the $beta$ function with two arguments. In the two-dimensional implementation the Sersic model has eight free parameters: $x_{cent}$, $y_{cent}$, $F_{tot}$, $r_{e}$, $n$, $c$, $q$, $PA$ We note that in place of fitting $\Sigma_{e}$, $F_{tot}$ is fitted instead, which is more often a useful parameter.

     The exponential profile and the total flux are simply

$$\Sigma(r)=\Sigma_{0}e^{-(r/r_{s})}$$

and

$$F_{tot}= 2 \pi r_{s}^{2}\Sigma_{0}q/R(c) $$

where $\Sigma_{0}$ is the central surface brightness and $r_{s}$ is the disk scale length. The relationship between the half-light radius $r_{e}$ and the scale length $r_{s}$ is $r_{s}$ = 1.678$r_{e}$ for this profile. Most disky galaxies are not composed of a single exponential disk, but also have either a central Sersic or de Vaucouleurs component. They may also have either a flat core or a truncated disk, which deviates from a simple exponential (e.g., van der Kruit 1979; Pohlen, Dettmar, \& Ltticke 2000). Two components (sersic bulge and exponential disc) and sometimes a third (point source) have been used to fit the surface distribution.

\begin{figure*}
\centerline{\includegraphics[width=20pc,height=20pc]{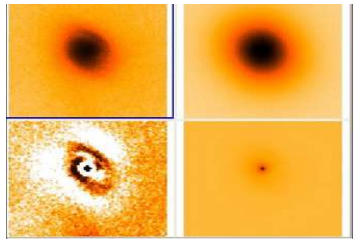}}
\caption{GALFIT example of a galaxy 0.34.s33.5485.1977 showing the $z$ band image, model, residual and the central point sources Note the spiral arms clearly visible in the residual.}
\label{page2}
\end{figure*}

The parameter $B/T$ is correlated with Hubble type, increasing for early type galaxies. The $\chi^{2}$ parameter is used to asses the quality of the fit, which is the difference between the model and the image, but in some cases (spiral galaxies, interacting galaxies with disturbed morphology) the goodness of the fit is assesed using the errors in fitting the parameters. As seen in Table \ref{morpho} the $B/T$ parameters have been obtained for the rest frame $B$ band and a neighbouring band, i.e. $B$ and $V$, $V$ and $i$, $i$ and $z$, and the values obtained are similar, even in the case of bad $\chi^{2}$ fits. Figure \ref{page2} shows an example of the GALFIT output for a galaxy at redshift 0.34, using its $z$ band image, model image, residual image and the central point source.

\subsection{Morphology labels}
By individually examining the galaxies, we have tried to assign a Hubble type to each galaxy. For the galaxies which are fit well, we divide them into five types in terms of the fraction of bulge luminosity in total: E/S0 (0.8 $<$ B/T $\leq$1), S0 (0.5 $<$ B/T $\leq$ 0.8), Sab (0.15 $<$ B/T $\leq$ 0.5), Sbc (0 $<$ B/T $\leq$ 0.15) and Sd (B/T = 0.5).
In Table \ref{morpho}, the galaxy type, quality factor (based on $\chi^{2}$ and errors in parameters) and interaction/merging type are listed. 

\small{
\begin{longtable}{llllllllll}
\label{morpho}

Band & Gal Id &            $z$   & $B/T^{g}$ & $\chi^{2}$ & $B/T$ & $\chi^{2}$ & $Type^{a,c}$ & $QF^{b}$ & HR\\
\hline
 \endfirsthead
\caption{continued.}\\
\hline
Band & Gal Id &            $z$   &$B/T$ &$\chi^{2}$& $B/T$ &$\chi^{2}$ & Type& QF& HR\\
\hline
\endhead
\hline
\endfoot
    &              &                 & $B$& ....  & $V$    & ....  &     &   &      \\
$B$ & 34.7178.7151 & 0.07,$0.60^{f}$ & 0.49 & 6.74   & ... &...  & I1  & 4 &  0.11 \\
$B$ & 24.8007.3700 & 0.07 & 0.07 & 119.84& ...  & ...    & SBc  & 1 & -0.70 \\
$B$ & 24.7958.4354 & 0.07 & 0.06 & 93.80 & 0.14 & 85.307 & SB    & 2 & -0.56\\
$B$ & 24.6423.6089 & 0.10 & 0.75 & 4.23 & 0.83  & 10.494 & E/S0  & 1  & -0.14\\
$B$ & 32.6833.6873 & 0.12 & 0.88 & 1.320 & 0.71 & 0.52   & E/S0  & 2  &  0.25\\
$B$ & 32.2961.4100 & 0.13 & 0.92 &  3.77 & ...  &  ...   & E (I2?)& 3  & -0.40                 \\
$B$ & 32.4959.3519 & 0.18 & 0.44 & 1.33 & 0.28  & 0.46   & $C^{d}$ & 2& -1.0 \\
$B$ & 34.286.4660  & 0.21 & 0.54 & 1.32 & 0.42  & 0.724  & SaB   & 1  & -0.69 \\
$B$ & 35.4274.1564 & 0.22 & 0.81 & 1.47 & 0.47  & 1.9  & E (M1)& 3  & -0.01\\
$B$ & 23.4055.829 & 0.24 & 1.00  & 0.14 & 0.56   & 9.46   & E (M1)& 3  & -1.0\\
$B$ & 32.6272.7584 & 0.24& 0.69  & 1.50 & 0.002  & 0.99   & S0 (I2)& 1      & -1.0 \\
$B$ & 12.3023.7881 & 0.25 & ...  &  ... &...     &  ...   & C      &  4    & -0.5\\
$B$ & 22.1647.5527 & 0.27 & 0.77 & 4.89 & 0.47  & 6.96   & E/S0(M1) & 3   & -1.0\\
$B$ & 23.7309.7934 & 0.29 & ...  &  ... &  0.83  & 0.503  &  C    &  4    & -1.0\\
$B$ & 43.1067.7094 & 0.30 & 0.65 & 1.28 & 0.57 & 1.21 & S0    & 1 & 0.08\\
$B$ & 33.5485.1977 & 0.34 & 0.67 & 1.86 & 0.82 & 2.86 & S0/I2 & 2 & -1.0\\
$B$ & 23.1569.4072 & 0.41 & 0.61 & 1.37 & 0.50 & 2.62 & S0    & 2 & -1.0 \\
$B$ & 33.2035.6494 & 0.36 & ... &...   & 0.01 & 2.18 & Sd    & 2 & -1.0\\\hline
   &              &                 &      $V$& ....  & $i$    & ....&     &   &      \\ \hline
$V$ & 44.86.7360   & 0.41 & 0.32 & 1.38 & 0.57 & 0.83 & Sab & 2 & -0.25\\
$V$ & 23.4072.5636 & 0.45 & 0.62 & 0.80 & 0.61 & 1.53 & I1 & 1 & -1.0\\
$V$ & 34.4164.2941 & 0.48 & 0.48 & 1.38 & 0.64 & 1.18 & C & 1  & 0.66 \\
$V$ & 34.5632.5597 & 0.52 & 0.41 & 1.8  & 0.5  & 568.9 & Sab & 1 & -1.0\\
$V$ & 35.2305.1442 & 0.53 & 0.84 & 1.12 & 0.91 & 1.26 & E/S0 & 1 & -0.47\\
$V$ & 43.1008.5558 & 0.54 & 0.51 & 299.19 & 0.53 & 125.4 & Sab(M2) & 3 & -0.52\\
$V$ & 44.1316.7197 & 0.54 & 0.45 & 192.50 & 0.54 & 1.235 & Sab & 2 & 0.16 \\
$V$ & 23.5223.1646 & 0.54 & 0.42 & 2.92 & ...   &  ...   & Sab (M1)& 2 & -1.0\\
$V$ & 34.5452.6404 & 0.56 & 0.15 & 22.39 & 0.3 & 2.57 & Sbc (M2)& 2 & -0.55\\
$V$ & 35.1984.2273 & 0.56,$0.73^{f}$ & 0.95 & 1.3 & 0.9 & 1.18 & E/S0 & 1 & 0.44 \\
$V$ & 44.3226.388 & 0.57 & 0.74 & 1.2 & 0.68 & 1.42 & S0 & 1 & 0.55\\
$V$ & 44.5670.2529 & 0.57& 0.59 & 1.74 & 0.58 & 3.75 & S0 & 1 & -1.0\\
$V$ & 33.7986.7039 & 0.57 & 0.61 & 1.4 & 0.91 & 1.18 & S0(I2) & 2 & -0.01\\
$V$ & 33.1644.1495 & 0.57 & 0.19 & 1.64 & 3.1e-06 & 1.817 &S0(I2) & 3 & -1.0\\
$V$ & 22.3771.7838 & 0.58 & 0.43 & 1.2 & 0.36 & 1.264 & Sab & 2 & -1.0\\
$V$ & 33.3974.5629 & 0.60 & 0.95 & 1.14 & ... &  ...  & E & 1 & 1.0\\
$V$ & 25.5367.2340 & 0.60 & 0.94 & 1.44 & 0.97 & 1.16 & E & 2 & \\
$V$ & 32.2115.509 & 0.61 & 0.87 & 1.72 & 0.98 & 1.47 & S0 & 2 & 1.0\\
$V$ & 13.3486.7640 & 0.62 & 1.00 & 8.04 & ... & ...  & E$^{e}$ & 2 & -0.21\\
$V$ & 25.4650.4320 & 0.66 & 0.60 & 0.5 & 0.97 & 1.23 & S0 & 1 & -0.49\\
$V$ & 22.703.432 & 0.66 & 0.77 & 1.28 & 0.92 & 1.45 &E/S0 & 1 & -0.17\\
$V$ & 35.814.827 & 0.66 & ...  & ...  &  0.49  & 1.2 & S0   &  4 & 0.52\\
$V$ & 24.4982.4300 & 0.66 & 0.70 & 1.1 & 0.86 & 1.459 &S0 & 1 & -0.40\\
$V$ & 12.8113.3010 & 0.66 & 0.68 & 1.47 & 0.38 & 1.4 & S0(I2) & 2 & 0.39\\
$V$ & 33.3306.6842 & 0.66 & 0.10 & 1.48 & 0.3 & 1.26 & Sbc & 1 & 1.0\\
$V$ & 24.6032.4845 & 0.66 & 0.94 & 1.1 & 0.88 & 1.19 & E & 1 & -0.14\\
$V$ & 44.2069.1289 & 0.66 & 0.77 & 277.34 & 0.69 & 1.53 & S0 & 2 & -1.0\\
$V$ & 33.4629.4857 & 0.67 & 0.38 & 1.13 & 0.41 & 1.22 & C & 3 & 0.03\\
$V$ & 32.6038.2074 & 0.67 & 0.48 & 24.12 & 0.49 & 171 & Sab & 2 & -1.0\\
$V$ & 32.6401.5996 & 0.67 & 0.98 & 1.34 & 0.95 & 1.425 & E & 1 & -0.45\\
$V$ & 34.7639.1510 & 0.67 & 0.94 & 1.1 & 0.60 & 1.48 & E & 1 & 0.34\\
$V$ & 44.553.6909 & 0.68 & 0.99 & 1.68 & 0.74 & 1.86 & E& 1 & \\
$V$ & 45.2709.1406 & 0.72& ...  &  ...  & 0.09 & 1.21 &  & 4 & 1.0\\
$V$ & 34.6908.6883& 0.73 & ...  & ...   & 0.38 & 1.48 &  & 4 & 1.0\\
$V$ & 35.1984.2273 & 0.73 & 0.95 & 1.12 & 0.91 & 1.19 & E & 1 & 0.44\\
$V$ & 33.3057.1046 & 0.73 & 0.88 & 1.47 & 0.81 & 1.20 & E & 1 & 1.0\\
$V$ & 35.1194.805 & 0.73 & 0.49 & 8.76 & 0.50 & 885.91 & Sab & 2 & 0.06\\
$V$ & 35.1087.2156 & 0.73 & 0.58 & 224.84 & 0.55 & 233.64 & Sab & 2 & -0.54\\
$V$ & 33.5063.6088 & 0.73 & 0.54 & 0.56 & 0.93 & 1.77 & M1 & 1 & -1.0\\
$V$ & 34.4872.651 & 0.73 & 0.85 & 1.23 & 0.83 & 1.21 & E & 1 & 1.0\\
$V$ & 24.7958.4354 & 0.73 & 0.14 & 85.31 & 0.11 & 93.71 & Sbc & 2 & -0.56\\
$V$ & 33.319.8027  & 0.74 & ...  &...    & ...  &  ...  & ... & 4 & -1.0 \\
$V$ & 23.2101.2233 & 0.74 & 0.53 & 1.18 & 0.57 & 1.34 & Sab & 2 & -0.51\\
$V$ & 33.3974.5629 & 0.76 & 0.95 & 1.24 & 0.95 & 1.16 & E & 1 & 1.0\\ \hline
   &              &                 &      $i$& ....  & $z$    & ....&     &   &      \\ \hline
$i$ & 35.1565.5405 & 0.78 &0.07 & 1.21 & ...& ...& Sbc & 2& -0.17 \\
$i$ & 23.6163.1447 & 0.83 & 0.90 & 1.42 & 0.61 & 1.24 & E & 1 & -1.0\\
$i$ & 25.4358.4829 & 0.83 & 0.34 & 114.77 & ...   & ...  & I1& 3 & -0.53 \\
$i$ & 21.5766.6718 & 0.83 & 0.41 & 1.38 & 0.87 & 1.25 & Sab(I2)/E & 2  & -0.32\\
$i$ & 24.6618.934  & 0.95 & ...  &  ... & ...  & ...  & I1        & 1  & -0.37\\
$i$ & 24.7927.1731 & 0.95 & 0.91    &  2.12    & 0.87 & 1.84 & I1 & 1  & -0.54\\
$i$ & 25.7031.484  & 0.96 & 0.37 & 1.48 & 0.46 & 1.19 & Sab & 1 & -0.23\\ \hline
   &              &                 &      $z$& ...   & $i$    & ... &     &   &      \\ \hline
$z$ & 34.6920.1658 & 0.96 & 0.18 & 0.42 & 0.92 & 1.15 &Sab & 2 & 1.0 \\
$z$ & 23.1562.1365 & 1.01& 0.64 & 1.15 & 0.75 & 1.18 & S0& 2 & 0.12\\
$z$ & 35.1314.3138 & 1.031& 0.52 & 12.96 &... &  ... & I1& 2 & -0.63 \\
$z$ & 33.7626.8017 & 1.03 &... & ...   &...& ...&... &... & 0.10 \\
$z$ & 44.4691.5854 & 1.03& 0.58 & 1.56 & 0.59 & 5.29 & Sab& 2 & -0.44\\
$z$ & 33.1991.439 & 1.03& 0.88 & 1.21 & ...   & ...  & E& 1 & 1.0\\
$z$ & 32.1926.7457 & 1.04 & 0.78 & 1.10 & 0.62 & 1.21 & S0(C)& 1 & 0.23\\
$z$ & 22.360.5312 & 1.06& 0.31 & 1.11 & 0.54 & 1.63 & Sab &2  & -1.0\\
$z$ & 32.5434.4203 & 1.09 & 0.06 & 1.22 & 0.03 & 1.34 & Sbc & 2 & 0.61\\
$z$ & 13.7935.1560 & 1.11 & 0.72 & 1.18 & 0.99 & 1.17 & I1 & 1& -1.0\\
$z$ & 13.4220.412 & 1.14 & 0.83 & 0.46 & 0.63 & 1.31 & E & 1& -1.0\\
$z$ & 24.419.4963 & 1.17 & 0.17 & 1.15 & 0.27 & 3.75 & Sbc & 2& 0.52\\
$z$ & 24.3934.455 & 1.21 & 0.57 & 11.10 & 0.51 & 13.88 & C & 2 & -0.20\\
$z$ & 12.6029.6234 & 1.22 & 0.50 & 1.25 & 0.5 & 6.51 & Sab & 1 & -0.45\\
$z$ & 24.7963.1519 & 1.22 & 0.73 & 1.55 & 0.54 & 3.08 & S0 & 2 & -0.47\\
$z$ & 34.2789.692 & 1.22 & 0.08 & 1.15 & 0.07 & 1.20 & Sbc & 1 & -1.0\\
$z$ & 23.5312.3027 & 1.30 & 0.43 & 1.17 & 0.69 & 1.23 & I1 & 1 & -1.0\\
$z$ & 22.8066.5328 & 1.31 & 0.21 & 1.16 & 0.15 & 1.23 & I1 & 1 & 0.60\\

\caption{Parameters of the optical counterparts of X-ray sources in HST/ACS GOODS-South field. The parameters are determined in their rest frame $B$ as well as the neighbouring band.}\\

\footnote{
Note to Table 1:\\
$^{a}$ Galaxy type -- E/S0: 0.8 $<$ $B/T$ $\leq$ 1, S0: 0.5 $<$ $B/T$~$\leq$ 0.8, Sab: 0.15~$<$ $B/T$~$\leq$ 0.5, Sbc: 0~$<$ $B/T$~$\leq$ 0.15, Sd: $B/T$~= 0, C: compact, T: tadpole, Irr: irregular. \\
$^{b}$ Quality factor -- 1: secure, 2: possibly secure, 3: insecure, 4: undetermined. \\
$^{c}$ Interaction/Merging -- M~1: obvious merging, M~2: possible merging, I1: obvious interaction, I2: possible interaction, R: relics of merger/interaction.  \\
$^{d}$ This object is too compact to obtain structural parameters.  \\
$^{e}$ Structural parameters are not successfully derived for this object from the two-dimensional structure fitting because of its clumpy light distribution and imagery close to the CCD chip border.\\
$^{f}$ More than one determination of spectroscopic redshift \\
$^{g}$ Structural parameters are not provided for the merging systems with multiple distinctly separated components.}

\end{longtable}

\subsection{Individual descriptions}
In Table 1. color maps of some of the objects are shown along with the $z$ band images. The images are in the order of increasing redshift and the identification is given in the form redshift.sector number.$x$coord.$y$coord of the GOODS ACS images. Complete data can be  obtained on request. A description of each galaxy is presented here.

{{\bf{0.07.34.7177.7151}}
 Possible interaction with a nearby elliptical galaxy (redshift not known), showing areas of massive star formation on the discs of both galaxies as well as two possible satellite dwarf galaxies.

{\bf{0.07.24.8007.3700}} Large Sb spiral inclined, color maps  show the entire disk as blue indicating star formation, brights clumps along possible bar.

{\bf{0.08.24.7958.4354}} Large disc galaxy showing many lots of star forming regions in the disc, making the disc very blue, the bulge seems redder.

{\bf{0.10.24.6422.6089}} Elliptical galaxy with two large blue arcs of star forming regions, probable relics of a merger or interaction

{\bf{0.12.32.6833.6872}} Disc galaxy with star formation taking place around the centre. Neighbouring object possibly interacting with the galaxy.

{\bf{0.13.32.2961.4100}} Disc galaxy with bright blue nucleus and star formation taking place on the entire disc. In possible interaction with a neighbouring spiral galaxy also showing star formation along its arms.

{\bf{0.18.32.4959.3518}} Compact blue galaxy.

{\bf{0.21.34.286.4660}} Disc galaxy with a bulge and red disc showing no signs of star formation, maybe reddened due to dust.

{\bf{0.22.35.4273.1564}} Elliptical galaxy with two nuclei or areas of star formation

{\bf{0.24.23.4055.829}} Disc galaxy with knots of star formation along the disc closer to a possibly interacting galaxy. The farther arm of the disc does not show star formation regions

{\bf{0.24.32.6272.7583}} Possible interacting pair of galaxies since active star formation seen in regions between the two.


{\bf{0.27.22.1647.5527}} Spiral galaxy with its disc clearly showing interaction with a nearby galaxy with disturbed morphology. An area of star formation is seen between the point of contact of the two galaxies.

{\bf{0.29.23.7309.7934}} Disc galaxy with a disturbed morphology, in possible interaction with its neighbouring galaxies

{\bf{0.30.43.1066.7094}} S0 with a prominent blue centre, disk reddened with probably dust features or old stellar population

{\bf{0.34.33.5485.1977}} Interacting pair of galaxies with clearly disturbed morphology. A single long arm is indicative of probable interaction between the galaxies

{\bf{0.36.33.2035.6494}} Disc galaxy with star formation taking place near the centre. Four patchy regions on the disc, possible areas of star formation, possibly due to past interactions or mergers

{\bf{0.41.23.1569.4072}} Disc galaxy with mild star formation taking place in knots along the disc or multiple nuclei due to past interactions.

{\bf{0.41.44.86.7359}} Sab with two nuclei not very blue probably due to low star formation or dust (?)

{\bf{0.45.23.4072.5635}} Interacting pair of galaxies with a large amount of star formation taking place in the nucleus of the larger galaxy and the whole disc of the smaller galaxy

{\bf{0.48.34.4164.2940}} Compact blue galaxy

{\bf{0.52.34.5632.5596}} Edge-on spiral galaxy(?), low star formation

{\bf{0.53.35.2304.1442}} Elliptical galaxy with highly reddened (dusty) nucleus

{\bf{0.54.43.1008.5558}} Sab with a very bright blue core and star formation taking place along a ring in the
galactic disc

{\bf{0.54.44.1316.7196}} Elliptical galaxy with some star formation taking place along two arcs on the disc

{\bf{0.54.23.5223.1646}}Elliptical galaxy with two distinct blue cores

{\bf{0.56.34.5452.6403}} Disc galaxy with a blue core and reddened disc

{\bf{0.56.35.1983.2272 }}Disc galaxy with star formation taking place at the centre as well as on the disc. The centre shows a very blue compact nucleus

{\bf{0.57.44.3226.388}} Disc galaxy with a prominent blue nucleus and a  disc

{\bf{0.57.44.5670.2529}} Disc galaxy with a prominent blue nucleus and a  disc

{\bf{0.57.33.7986.7039}} Reddened galaxy with a star forming ejection/arm possibly caused due to interaction with neighbouring two objects

{\bf{0.57.33.1643.1495}} S0 galaxy with star formation taking place in two arcs along the disc, probably due to some past interaction

{\bf{0.58.22.3770.7838}} Sab galaxy, quite compact and hence not showing many features

{\bf{0.60.33.3973.5629}} Blue star forming bulge surrounded by a disc.

{\bf{0.60.25.5366.2340}} Reddened disc with no signs of star formation in the galaxy

{\bf{0.61.32.2115.509}} S0 galaxy with a blue core

{\bf{0.62.13.3485.7640}}Very poor S/N image as it is close to the edge

{\bf{0.66.25.4650.4319}} Galaxy with a blue core and reddened disc

{\bf{0.66.22.703.431}} Galaxy with a blue core and reddened disc

{\bf{0.66.35.813.826}} Galaxy with reddened disc

{\bf{0.66.24.4981.4299}} Galaxy with reddened disc

{\bf{0.66.12.8113.3009}} Galaxy with a single arm probably formed by some interaction

{\bf{0.66.33.3306.6842}}Spiral(?) galaxy close to the edge of the frame

{\bf{0.66.24.6031.4845}} Galaxy with reddened disc

{\bf{0.66.44.2069.1289}} Galaxy with a blue core and reddened disc

{\bf{0.67.33.4629.4857}} Compact red galaxy

{\bf{0.67.32.6038.2074}} Galaxy with reddened disc with a little bit of star formation on the disc

{\bf{0.67.32.6401.5996}} Galaxy with reddened disc and a little bit of star formation in a ring around the nucleus

{\bf{0.67.34.7638.1510}} Elliptical galaxy with a read bulge, no signs of present star formation.

{\bf{0.68.44.553.6909}} Galaxy with a blue core and reddened disc

{\bf{0.72.45.2709.1406}}

{\bf{0.73.34.6908.6883}} The galaxy looks like  a barred spiral of type Sbc, with a reddened bulge, probably in interaction with a nearby dwarf galaxy in direct contact with it showing intense star formation on the whole of the dwarf and the part of the disc of the galaxy in contact too.

{\bf{0.73.35.1984.2272}} Elliptical galaxy with some star formation taking place near the core

{\bf{0.73.33.3056.1045}}  Elliptical galaxy with very blue compact core.

{\bf{0.73.35.1194.804}} Elliptical galaxy with star formation along two arc in the disc

{\bf{0.73.35.1087.2156}} Disc galaxy with star formation taking place almost throughout the galaxy and more enhanced on the core and a ring on the outerpart of the disc

{\bf{0.73.33.5063.6088}} Blue galaxy with a small satellite galaxy(?) both showing high star formation activity. The galaxy has a clear bulge and exponential disc

{\bf{0.73.34.4872.651}} Galaxy with a blue bulge and a reddened disc

{\bf{0.73.24.7858.2362}} Galaxy with star formation taking place on the central core and a ring along the disc

{\bf{0.73.33.319.8027}} Elliptical galaxy interacting with two nearby galaxies with star formation taking place in the disc as well as the nearby galaxy

{\bf{0.74.23.2100.2233}} Galaxy with a red bulge and a very blue starforming disc

{\bf{0.76.33.3974.5629}} Elliptical galaxy probably interacting with a nearby faint galaxy. Star formation taking place in the centre while the disk is quite reddened (probably due to dust)

{\bf{0.78.35.1565.5405}} Disc galaxy with a slightly asymmetric disc, single arm probably due to some interaction.

{\bf{0.83.23.6162.1446}} Blue compact disc galaxy with strong star formation taking place

{\bf{0.83.25.4358.4829}} Galaxy probably in interaction with neighbouring  galaxy

{\bf{0.83.21.5766.6718}} Sb type spiral galaxy, with one arm longer due to possible interaction with neighbouring fuzzy object. Active star formation on the longer arm and the neighbouring blue object indicate on induced star  formation due to possible interaction.

{\bf{0.95.24.6618.933}} Spiral galaxy with an elongated arm probably due to interaction with the other galaxies in the neighbourhood. Color maps confirm interaction

{\bf{0.95.24.7927.1731}} Galaxy seems part of a cluster. Seems to be a direction connection to a nearby object, and has a blue compact nucleus with star formation in many of the nearby galaxies

{\bf{0.95.25.7031.484}} Spiral galaxy in close interaction with a neighbouring galaxy, star formation taking place along the complete bridge between the two galaxies.

{\bf{0.96.34.6920.1658}} Red bulge, but star formation taking place along an arm (spiral or caused by interactions)

{\bf{1.01.23.1562.1365}} Heavily reddened compact galaxy

{\bf{1.031.35.1314.3138}} Spiral Sb galaxy (or long arm created out of interaction) with a nearby object also showing star formation along the disc closer to the galaxy which shows star formation on the complete disc


{\bf{1.037.44.4690.5853}}Compact galaxy showing star formation along the complete disc

{\bf{1.037.33.1991.438}} Compact blue elliptical galaxy with mild star formation along the disc

{\bf{1.043.32.1925.7457}} Compact elliptical galaxy with star formation along the disc

{\bf{1.06.22.360.5312}} Red bulge with star formation taking place along a ring on the disc

{\bf{1.09.32.5433.4203}} Elliptical galaxy with probably lots of reddening due to dust or no star formation along the disc

{\bf{1.11.13.7935.1560}} Elliptical galaxy with no star formation taking place

{\bf{1.14.13.4220.412}} Elliptical galaxy with highly reddened bulge and disc. Probabale dust features visible in residual image.

{\bf{1.17.24.419.4963}} Elliptical galaxy with possible interaction with neighbouring two galaxies

{\bf{1.21.24.3934.455}} Compact blue galaxy

{\bf{1.22.12.6029.6234}} Elliptical galaxy with star formation taking place along a ring on the disc.

{\bf{1.22.24.7962.1519}} Compact galaxy with star formation taking place on the disc

{\bf{1.22.34.2789.692}} Spiral galaxy with some star formation taking place along the arms

{\bf{1.30.23.5312.3027}} Galaxy with a long arm (arc) along the disc showing active star formation

{\bf{1.31.22.8066.5328}} Probable spiral Sc galaxy with active star formation along the complete disc

\section{Color Maps}
To obtain color maps, we used the $B-z$ images. The ACS/GOODS images are astrometrically aligned by the GOODS team with an accuracy of better than a fraction of a pixel. The color maps are obtained by dividing the $B$ band image by the $z$ band image after correcting them by their appropriate zeropoints.
The color maps and images in the $z$ band for the galaxies are shown in Table 1 (complete table in the paper). The method adopted to obtain color maps has been described in Zheng at al (2004b). The advantage being that the color maps very well show regions of star  formation, dust, etc. In Table 1, the color bar next to the color map shows the color scheme in the observed frame. The color range is adjusted for best visualization. Each image is labelled where the first number is the redshift while the id is created by sector number.xsector.ysector to identify it in the GOODS ACS images.

\begin{tabular}{llll}

Image & Colormap & Image & Colormap\\

0.07(0.6).34.7178.7151& &0.08.24.8007.3700&\\
\includegraphics[width= 0.20\linewidth]{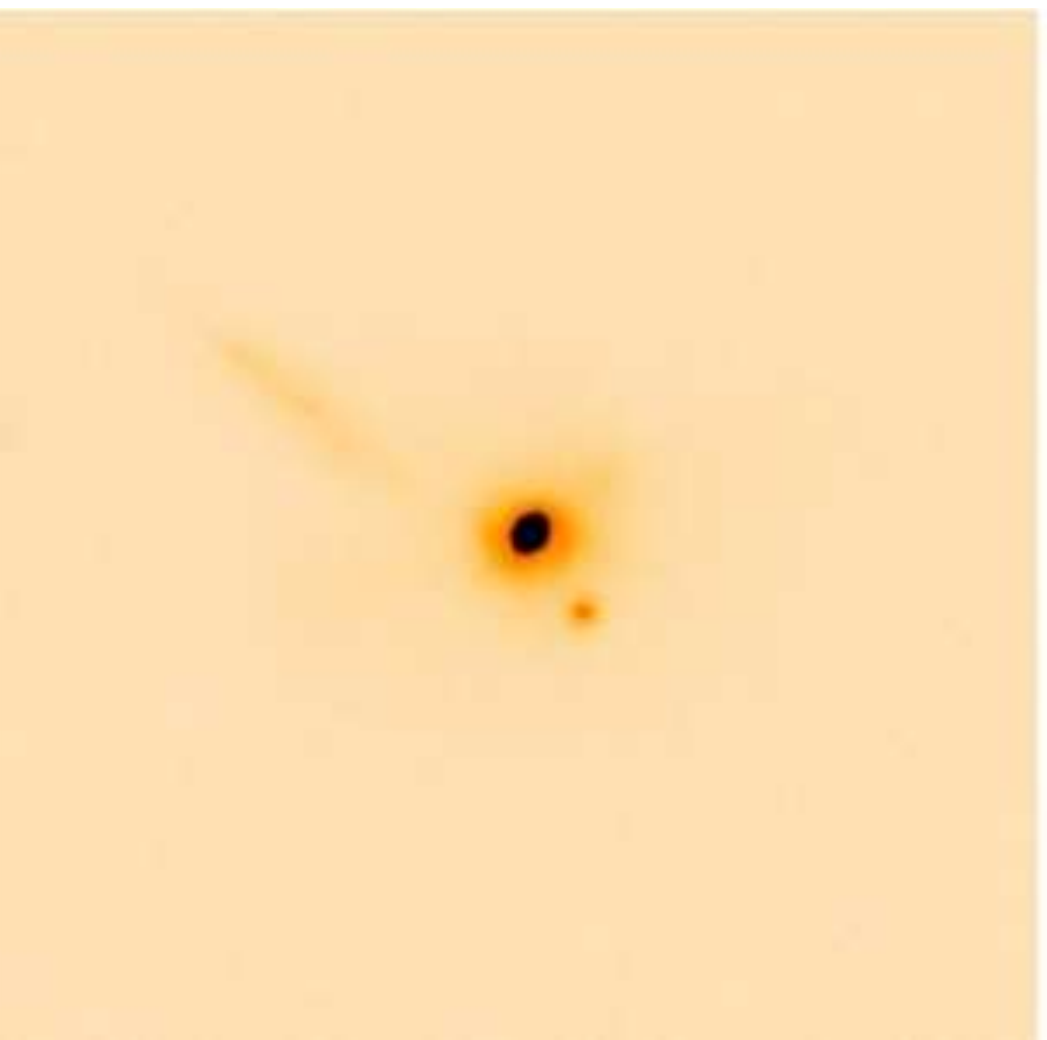} &\includegraphics[width= 0.23\linewidth]{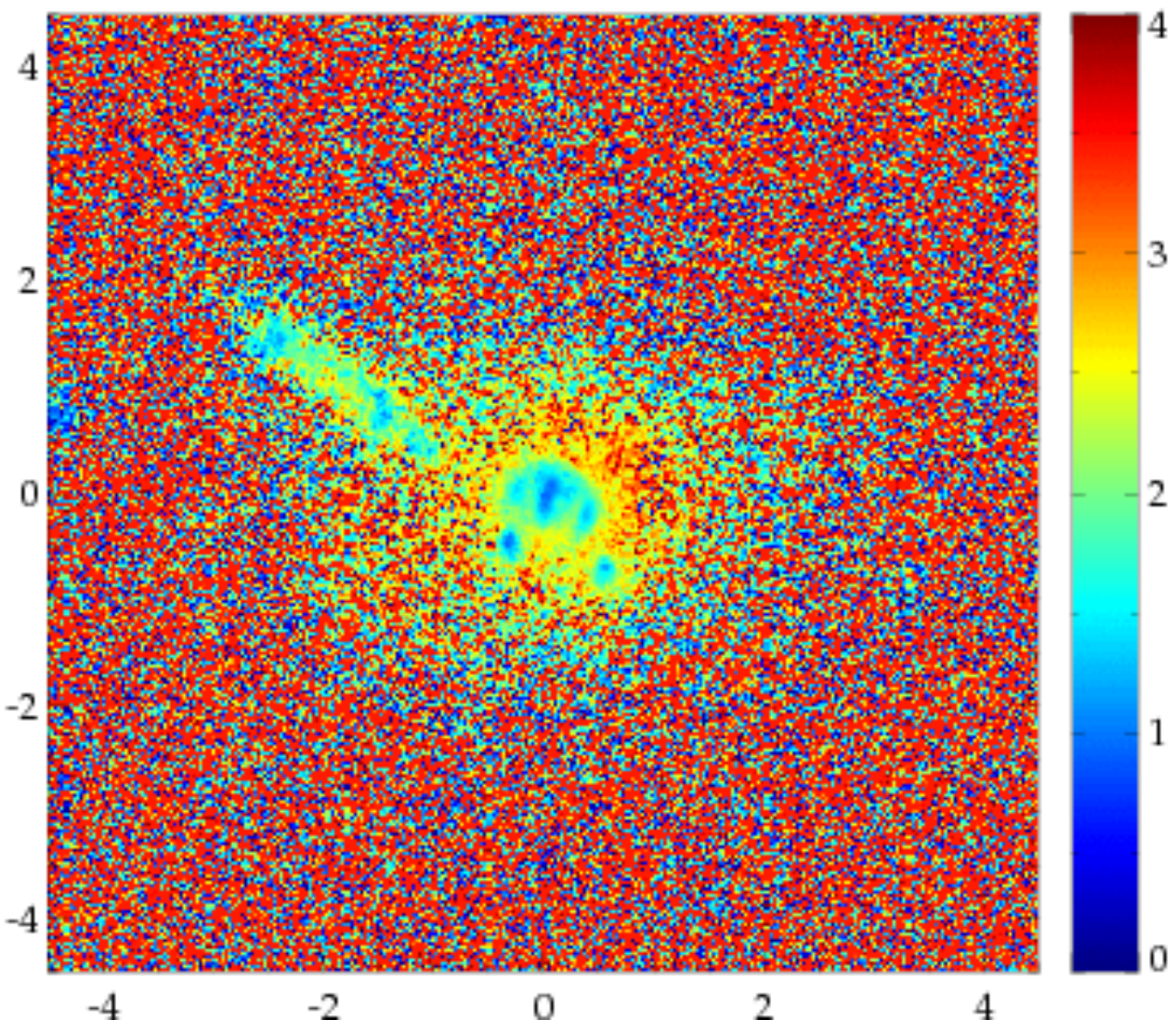} &\includegraphics[width= 0.20\linewidth]{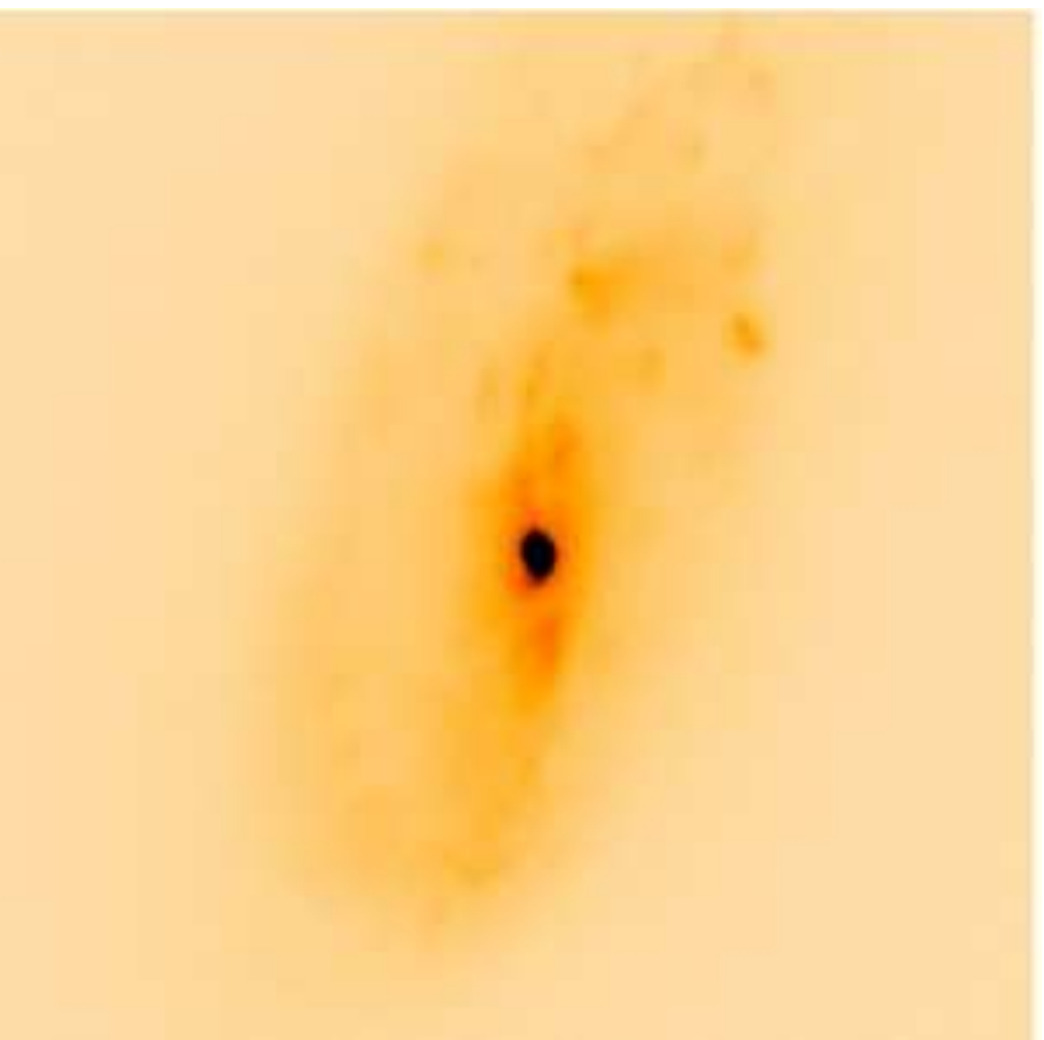} &\includegraphics[width= 0.23\linewidth]{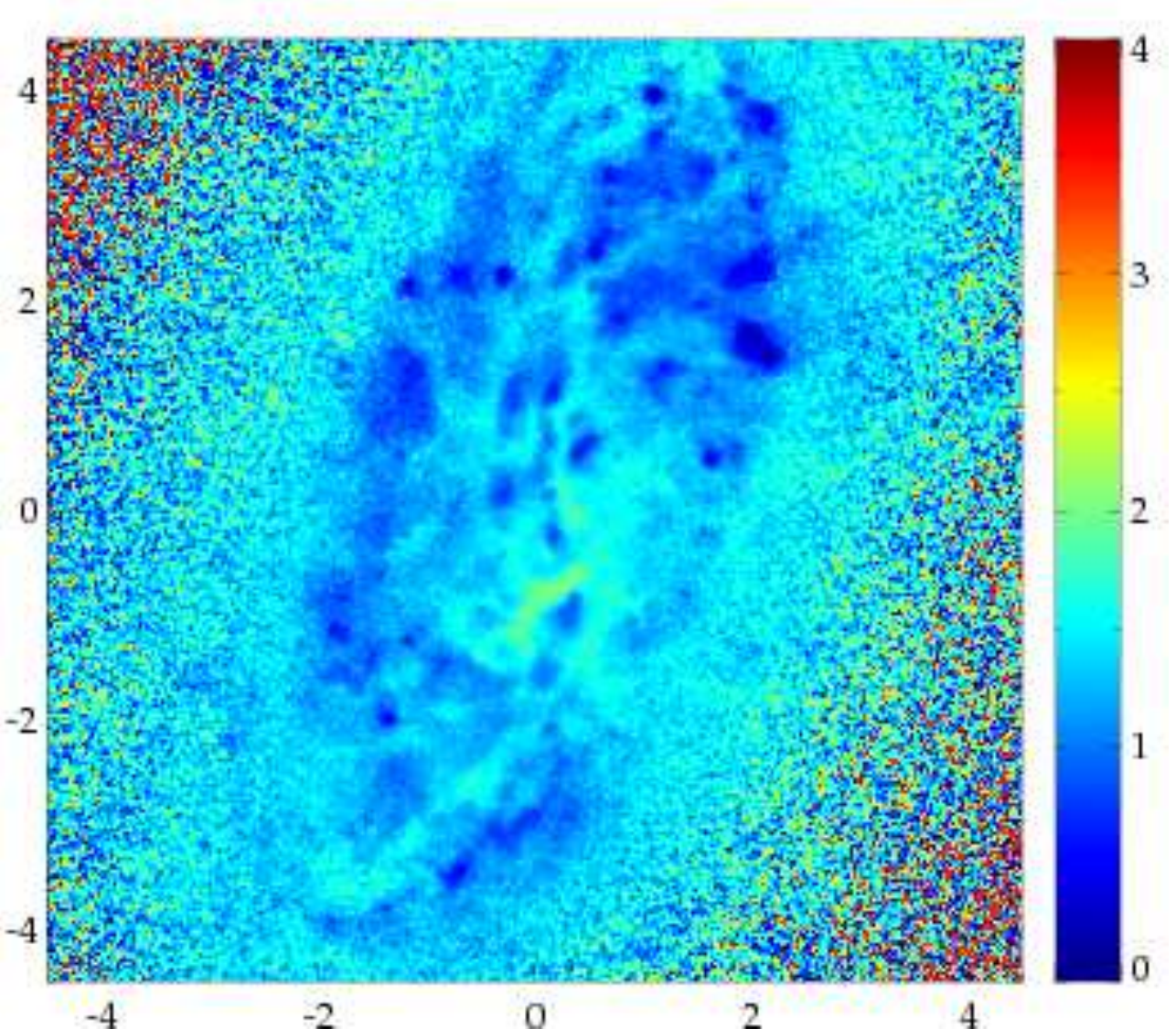} \\

0.08.s24.7958.4354&&0.10.s24.6423.6089&\\
\includegraphics[width= 0.20\linewidth]{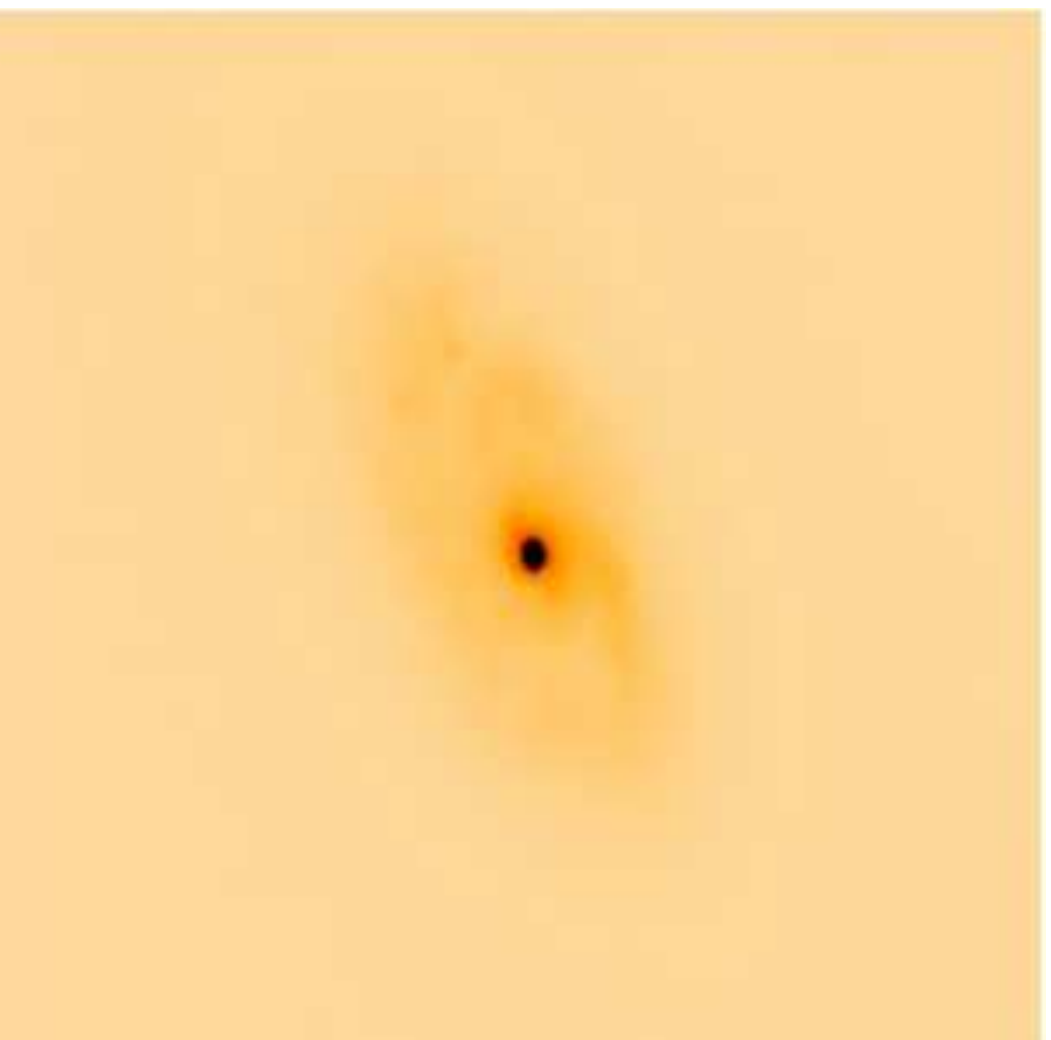} &\includegraphics[width= 0.23\linewidth]{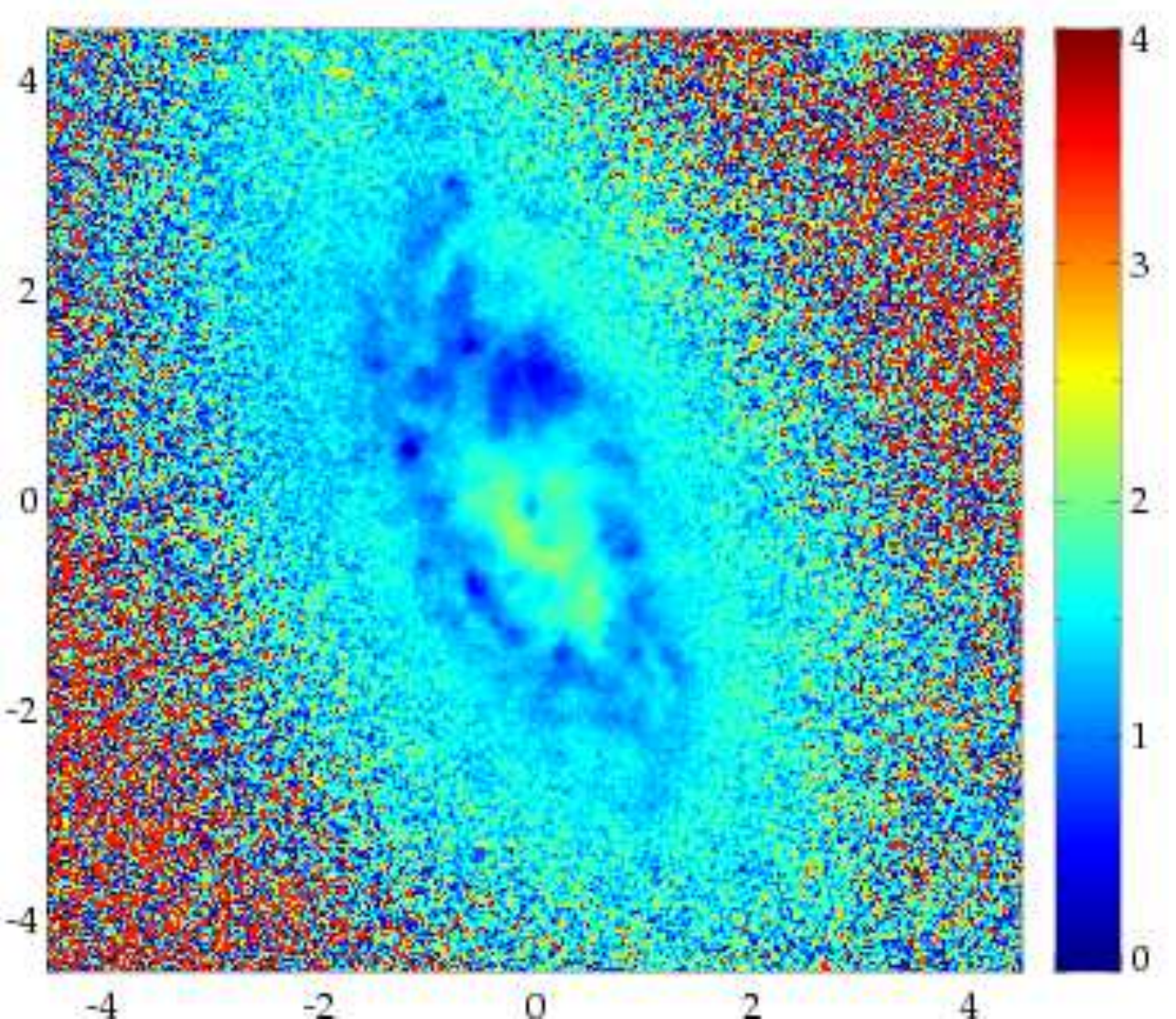}  &\includegraphics[width= 0.20\linewidth]{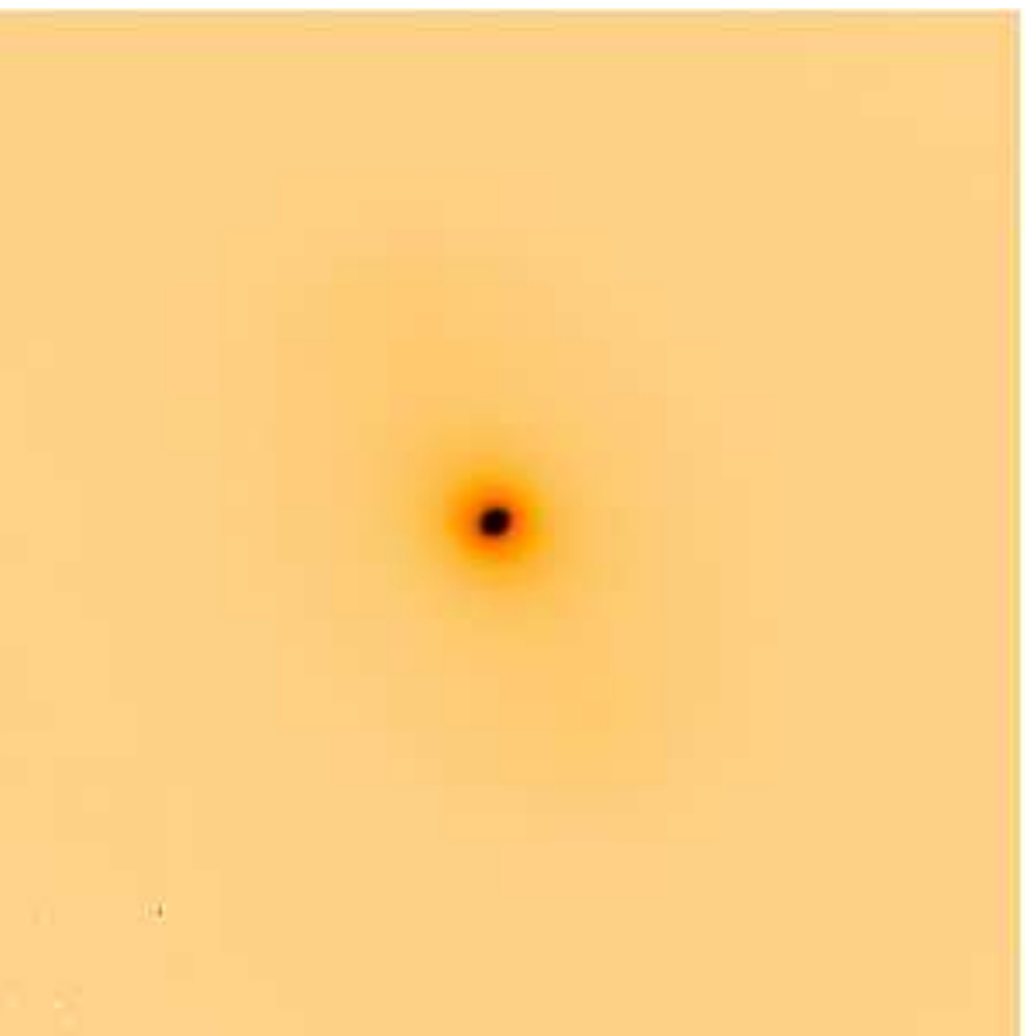} &\includegraphics[width= 0.23\linewidth]{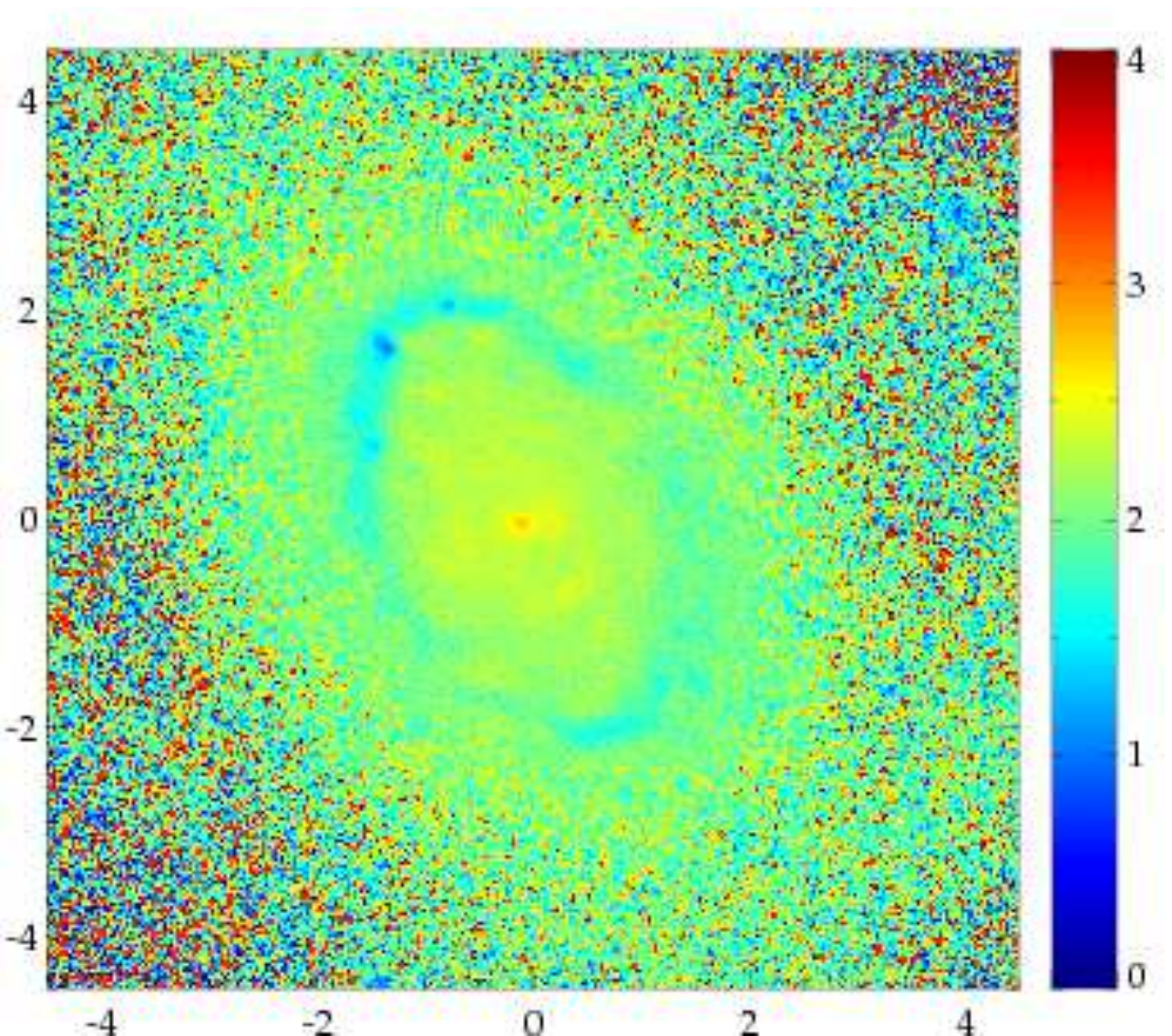} \\

\label{colormaps}
\end{tabular}
`
\section{Results and Conclusions}
Using HST/ACS images in four bands F435W, F606W, F775W and F850LP, we identify optical counterparts to the X-ray sources in the Chandra Deep Field South in the GOODS South field. A detailed study has been made of these sources to study their morphological types. Figure \ref{morphoz} shows the spread of morphological types for the different redshift bins. Figure ~\ref{dbz} shows the change in Bulge/Total ratios of the galaxies at different redshift bins. Decomposition of the galaxy luminosity profiles has not been an easy task. For many galaxies, the $\chi^{2}$ obtained from galfit differs strongly from 1. In Fig. \ref{dbz}, the solid lines indicate galaxies with good  $\chi^{2}$ fits (i.e. $\chi^{2} \approx 1$), while the dotted lines include those with bad $\chi^{2}$ (i.e.  $\chi^{2} \geq 1.4$). It can be seen from the figure that in both cases, the trend towards more bulge dominated galaxies is obvious.

To compare the AGN population with the non-AGN population, we use the HST/ACS morphologies from Bundy, Ellis and Conselice (2005). Using a well-defined catalog of 2150 galaxies in the GOODS field, they have determined morphological types of a large number of galaxies in the GOODS fields. The catalog is  publically available  (www.astro.caltech.edu/GOODS\_morphs/). As the number of Chandra sources is about 1\% of the total galaxies, we use their catalog to describe the general population of galaxies, as the number of AGN hosts in their sample (if they exist) is very small. Table \ref{bundy} shows the data used from their catalog (All) as well as that obtained by us for our sample (AGN).

\begin{table}[h]
\small
\begin{tabular}{|l|ll|ll|ll|}
\hline
Morpho Type    & $0.2 \leq z$ & $\leq 0.51$ & $0.51 \leq z$ & $\leq 0.8$ & $0.8 \leq z$ & $\leq 1.4$\\
\hline
               & AGN & All & AGN & All & AGN & All\\
 Spirals       & 2   & 353     & 10  & 326     & 9   & 291\\
 Ellipticals   & 3   & 167     & 21  & 220     & 6   & 131\\
 Interacting   & 6   & 126     & 8   & 138     & 7   & 154\\
 
\hline
\end{tabular}
\caption[]{Distribution of morphological types for AGN and non-AGN (Bundy et al 2005) }
\label{bundy}
\end{table}

 At low redshifts ($z\leq 0.4$), we are biased due to small numbers as well as selection effects and the majority of AGN host galaxies are interacting, while in the non-AGN population, spirals dominate. For the redshift bin $0.51 \leq z \leq 0.8$, bulge-dominated early type galaxies are the majority in the AGN population,  while for the non-AGNs spirals continue to dominate. For redshifts $\geq$ 0.6, the percentage of interacting galaxies in the AGN and non-AGN population do not differ significantly.
 
 For the data from Bundy et al (2005), we performed  $\chi^{2}$ test for the distribution of morphological types (spirals, ellipticals and interacting) in different redshift bins ($0.2 \leq z \leq 0.551, 0.51 \leq z \leq 0.8$ and $0.8 \leq z \leq 1.4$) and found  $\chi^{2}= 22.8085$ with 4 degrees of freedom and $p =0.0001$ \footnote{$p \leq 0.05$ implies statistical significance to 95 \%}; implying a very strong statistical significance. This indicates that the three morphological types are significantly different with respect to redshift.

We did a similar $\chi^{2}$ test for our sample in the same redshift bins and obtained no statistical significance ($\chi^{2}$ =8.1742, p=0.0854). This could be because of the small numbers in the sample. However, when we did a $\chi^{2}$ test on the summed up numbers of spirals and interacting galaxies with the ellipticals for $z \geq 0.51$, we obtained a $\chi^{2}=4.020$ with one degree of freedom and $p$=0.0448. This is important since it shows that the apparent decrease in in the E/S0 fraction and increase in spiral and merging classes is significant as we go from $z \geq 0.51$. This is consistent with the heirarchial merging and nuclear feeding hypothesis.

\begin{figure*}
\centerline{\includegraphics[width=20pc,height=20pc]{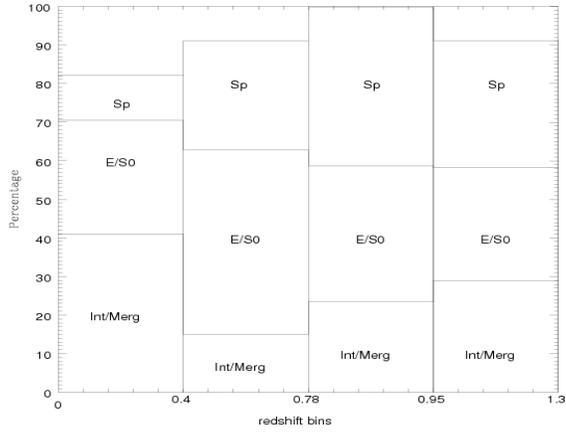}}
\caption{Distribution of different morphological types with redshift}
\label{morphoz}
\end{figure*}

\begin{figure*}
\centerline{\includegraphics[width=20pc,height=20pc]{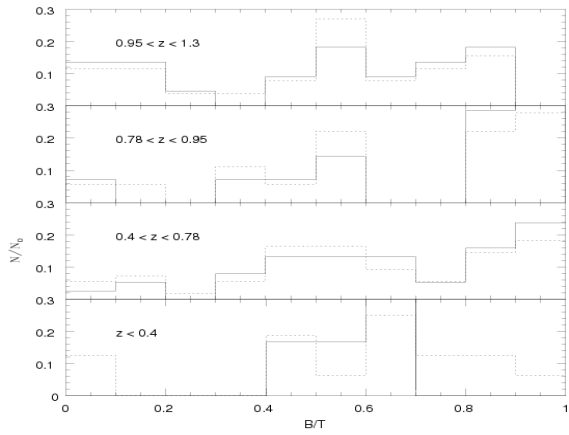}}
\caption{Evolution of the Bulge/Total ratio with redshift}
\label{dbz}
\end{figure*}


Grogin et al (2005) came to a similar conclusion using CAS parameters. They indicate that for moderate luminosity AGNs, mergers are not the key fueling source. The AGN is probably fueled by disc instabilities or other processes. They concluded that AGN host galaxies clearly show a bulge dominated population, though the incidence of merging activity is not very obvious. 

Another point to be noted is that the CDFS has two prominent spikes in the redshift distribution at $z=0.67$ and $z=0.73$, this region contains almost 1/3 of the identified sources. And there does exist a prominent overdensity of AGN in this region compared to the non-AGNs. Hence one should not rule out the possibilty of some influence of clustering on the AGNs and the morphology of their hosts.
 
 Early studies by Bahcall et al (1997) for a sample of 20 nearby ($z \leq 0.3$) quasars indicated that the host galaxies of highest luminosity AGNs ($L_X \geq 10^{44.5}$ egs s$^{-1}$) are mostly interacting.


The results presented in this paper apply to moderate luminosity AGN ($L_X \leq 10^{43.5}$ egs s$^{-1}$) at redshifts $z\approx 0.4-1.3$. The GOODs survey covers a very small solid angle  and thus  very few high luminosity AGN are present in this sample to test the results obtained by Bahcall et al (1997) over the same redshift range. A much more wider survey was the Galaxy Evolution from Morphology and SEDs (GEMS) project  where Sanchez et al (2004) studied the host galaxies of 15 optically selected AGN with $ 0.5\leq z \leq 1.1 $. Sanchez et al (2004) found that 80\% of the host galaxies have early-type (bulge-dominated) morphologies, while the rest have structures characteristic of late-type (disk-dominated) galaxies. They found that 25\% of the early types and 30\% of the late types exhibit disturbances consistent with galaxy interactions. This again indicates the AGN hosts are surely  bulge dominated, but not dominated by merger activity. At present, however, the samples are still very small and future studies with more high luminosity AGN will be required to shed more light on this trend.

\section{acknowledgements}
       The author is very thankful to the referee for valuable comments and suggestions. The author is also very grateful to Ajit Kembhavi, Abhishek Rawat, S. N. Hasan, Francois Hammer, Hector Flores, Xianzhong Zheng, Haragopal Vajjha and  Swara Ravindranath for discussions and help. Spectropscopic redshifts were obtained using observations have been carried out using the Very Large Telescope at the ESO Paranal Observatory under Program ID: LP170.A-0788. This work was funded by Project No  2804-1 of the Indo-French Centre for the Promotion of Advanced Research (CEFIPRA).

\end{document}